\documentclass[aps,prb,twocolumn,superscriptaddress,amsmath,showpacs]{revtex4-1}
\usepackage{graphicx}
\usepackage{float}

\begin{document}


\title{Isotropic and Anisotropic Regimes of the Field-Dependent Spin Dynamics in Sr$_{2}$IrO$_{4}$:  Raman Scattering Studies }


\author{Y. Gim}
\affiliation{Department of Physics and Frederick Seitz Materials Research Laboratory, University of Illinois, Urbana, Illinois 61801, USA}
\author{A. Sethi}
\affiliation{Department of Physics and Frederick Seitz Materials Research Laboratory, University of Illinois, Urbana, Illinois 61801, USA}
\author{Q. Zhao}
\affiliation{Material Science Division, Argonne National Laboratory, Argonne, Illinois 60439, USA}
\author{J. F. Mitchell}
\affiliation{Material Science Division, Argonne National Laboratory, Argonne, Illinois 60439, USA}
\author{G. Cao}
\affiliation{Center for Advanced Materials, University of Kentucky, Lexington, Kentucky 40506, USA}
\affiliation{Department of Physics and Astronomy, University of Kentucky, Lexington, Kentucky, 40506, USA}
\author{S. L. Cooper}
\affiliation{Department of Physics and Frederick Seitz Materials Research Laboratory, University of Illinois, Urbana, Illinois 61801, USA}



\date{\today}

\begin{abstract}
A major focus of experimental interest in Sr$_{2}$IrO$_{4}$ has been to clarify how the magnetic excitations of this strongly spin-orbit coupled system differ from the predictions of an isotropic 2D spin-1/2 Heisenberg model and to explore the extent to which strong spin-orbit coupling affects the magnetic properties of iridates.  Here, we present a high-resolution inelastic light (Raman) scattering study of the low energy magnetic excitation spectrum of Sr$_{2}$IrO$_{4}$ and doped Eu-doped Sr$_{2}$IrO$_{4}$ as functions of both temperature and applied magnetic field.  We show that the high-field (H$>$1.5 T) in-plane spin dynamics of Sr$_{2}$IrO$_{4}$ are isotropic and governed by the interplay between the applied field and the small in-plane ferromagnetic spin components induced by the Dzyaloshinskii-Moriya interaction. However, the spin dynamics of Sr$_{2}$IrO$_{4}$ at lower fields (H$<$1.5 T) exhibit important effects associated with interlayer coupling and in-plane anisotropy, including a spin-flop transition at H$_{c}$ in Sr$_{2}$IrO$_{4}$ that occurs either discontinuously or via a continuous rotation of the spins, depending upon the in-plane orientation of the applied field.  These results show that in-plane anisotropy and interlayer coupling effects play important roles in the low-field magnetic and dynamical properties of Sr$_{2}$IrO$_{4}$.
\end{abstract}


\maketitle

\section{Introduction}
The 5d iridates have attracted much recent attention, because the comparable spin-orbit coupling and electronic correlation energy scales in these materials are expected to be conducive to exotic phases and phenomena, such as superconductivity,\cite{Wang2011, You2012} spin-liquid states,\cite{Okamoto2007} a J$_{eff}$=1/2 Mott state,\cite{Kim2008,Kim2009} topological phases,\cite{Pesin2010,Shitade2009,Kim2012c} and magneto-electric behavior.\cite{Chikara2009}  The antiferromagnetic insulator, Sr$_{2}$IrO$_{4}$ (T$_{N}$$\sim$240 K),\cite{Cao1998,Ge2011,Ye2013} is of particular interest, because the strong spin-orbit coupling in this iridate contributes significantly to its insulating behavior:\cite{Kim2008,Kim2009}  The Ir$^{4+}$ (5d$^{5}$) ions in Sr$_{2}$IrO$_{4}$ provide 5 d-electrons that occupy the t$_{2g}$ states, which are well separated from higher-energy e$_{g}$ bands by crystal-field effects.  The spin-orbit interaction $\lambda\sim$0.4 eV further splits the t$_{2g}$ levels into half-filled J$_{eff}$=1/2 and filled J$_{eff}$=3/2 bands.  Insulating behavior is thought to arise because of on-site Coulomb interactions, which split the J$_{eff}$=1/2 level into upper and lower Hubbard bands.  This general picture is supported by angle-resolved photoemission, optical conductivity, and x-ray absorption measurements.\cite{Moon2006,Kim2008}

An important unresolved issue concerns the extent to which the strongly entangled spin-orbit states comprising the J$_{eff}$=1/2 Ir 5d moments influence both the magnetic properties and the magnetic excitation spectrum of Sr$_{2}$IrO$_{4}$.\cite{Kim2008,Kim2009,Kim2012b}  Neutron scattering studies \cite{Crawford1994,Ye2013} have shown that Sr$_{2}$IrO$_{4}$ has an antiferromagnetic configuration with a $\sim$$11^{\circ}$ canting of the AF spins in the ab-plane, associated with the antisymmetric Dzyaloshinskii-Moriya (DM) spin exchange anisotropy in Sr$_{2}$IrO$_{4}$.  This in-plane canting leads to net ferromagnetic in-plane moments (see Fig. \ref{Fig1}(c)), m$_{FM} \approx$0.06$\mu_{B}$,\cite{Fujiyama2012,Bahr2014} that are antiferromagnetically coupled along the c-axis in a ($\uparrow\downarrow\downarrow\uparrow$) pattern.\cite{Kim2009}  

However, while the static magnetic configuration of Sr$_{2}$IrO$_{4}$ has been well established, the magnetic excitation spectrum of this strongly spin-orbit-coupled system has not been so well characterized.  In particular, it is not clear whether the magnetic excitations associated with the spin-orbit entangled J$_{eff}$=1/2 moments in Sr$_{2}$IrO$_{4}$ can be described by the predictions of an isotropic S=1/2 Heisenberg model.\cite{Fujiyama2012,Igarashi2013,Igarashi2014}  Unfortunately, inelastic neutron scattering studies of the magnetic excitations in Sr$_{2}$IrO$_{4}$ are hampered by the strong absorption of neutrons by Ir.\cite{Igarashi2014}  Resonant inelastic x-ray scattering (RIXS) studies of Sr$_{2}$IrO$_{4}$ have probed both low-energy charge \cite{Ishii2011} and magnetic \cite{Kim2012b,Liu2015} excitations, but the relatively low resolution associated with RIXS measurements have not allowed a detailed study of the low energy magnetic excitations that would reveal deviations from Heisenberg model predictions.  Interestingly, recent resonant magnetic diffuse x-ray \cite{Fujiyama2012} and field-dependent electron spin resonance \cite{Bahr2014} studies of Sr$_{2}$IrO$_{4}$ have offered evidence that the magnetic correlations and excitations are well-described by the two-dimensional S=1/2 Heisenberg model, in spite of the strong spin-orbital coupling associated with the Ir 5d moments. 

In this paper, we present an inelastic light (Raman) scattering study of the low energy magnetic excitation spectrum of Sr$_{2}$IrO$_{4}$ and doped Eu-doped Sr$_{2}$IrO$_{4}$ as functions of temperature, applied magnetic field, and magnetic field orientation. Inelastic light scattering is a valuable probe for studying the spin-dynamics of Sr$_{2}$IrO$_{4}$:  This technique is a very high resolution probe of the \textbf{q}=0 magnetic excitation energies, which are influenced by small anisotropy and interlayer coupling interactions that can uncover physics beyond the isotropic, two-dimensional S=1/2 Heisenberg model description.\cite{Chovan2000, Gozar2004,Benfatto2006a,Benfatto2006b}  Additionally, Raman scattering can probe the spin dynamics both with and without an applied field. Consequently, this technique is useful for studies of spin dynamics in the interesting low-field region of Sr$_{2}$IrO$_{4}$, particularly through the field-induced antiferromagnetic (AF) to weakly ferromagnetic (WFM) spin-flop transition at H$_{c}$$\approx$0.15 T.\cite{Cao1998,Kim2009}

In the results reported here, we show that the in-plane spin dynamics of Sr$_{2}$IrO$_{4}$ at high fields (H$>$1.5 T) are well-described by isotropic, two-dimensional S=1/2 Heisenberg model predictions. By contrast, the low-field (H$<$1.5 T) spin dynamics of Sr$_{2}$IrO$_{4}$ exhibit important effects associated with interlayer coupling and in-plane anisotropy that are not accounted for in standard descriptions of the spin-dynamics of Sr$_{2}$IrO$_{4}$. These effects include an anisotropic field-dependence of the spin-dynamics for H$<$1.5 T, and an AF-to-WFM transition that occurs via either discontinuous spin-flop or continuous spin-reorientation transitions for different in-plane field orientations.

\section{Experimental Set up}
\subsection{Sample Preparation}
The single crystals of Sr$_{2}$IrO$_{4}$ (T$_{N}$$\sim$240K) studied were grown from off-stoichiometric quantities of SrCl$_{2}$, SrCO$_{3}$ and IrO$_{2}$ using self-flux techniques. Technical details are described elsewhere.\cite{Cao1998}  The structures of Sr$_{2}$IrO$_{4}$ samples were determined using a Nonius Kappa CCD X-ray diffractometer. The data were collected between 90 K and 300 K, and the structures were refined using the SHELX-97 program.\cite{Sheldrick2008} Chemical compositions of the single crystals were determined using energy dispersive X-ray analysis (EDX) (Hitachi/Oxford 3000).

The Eu-doped Sr$_{2}$IrO$_{4}$ sample (T$_{N}$$\sim$200K) was synthesized at Argonne National Laboratory using a Eu-enriched SrCl$_{2}$ flux method. Samples were characterized by DC magnetization using a Quantum Design SQUID magnetometer.\footnote{STM measurements on the Eu-doped sample (A. Satpathy,  private communication) indicate that the Eu concentration is $<$1\% and suggests the possible role of O vacancies at the $\sim$1\% level. Both species in principle are electron donors}

The samples were cleaved to create c-axis normal surfaces, as verified using room temperature x-ray diffraction measurements.

\subsection{Raman Measurements}
Raman scattering measurements were performed using the 647.1 nm excitation line from a Kr$^{+}$ laser. The incident laser power was limited to 5 mW and was focused to a $\sim$50 $\mu$m-diameter spot to minimize laser heating of the samples. The scattered light from the samples was collected in a backscattering geometry, dispersed through a triple stage spectrometer, and then detected with a liquid-nitrogen-cooled CCD detector.  The incident light polarization was selected with a combination of a polarization rotator and a 1/4-waveplate and the scattered light polarization was analyzed with a linear polarizer. The scattering geometry used for all measurements had both the incident and scattered polarizations oriented in the ab-planes of the crystals. The incident and scattered light polarizations, \textbf{e}$_{i}$ and \textbf{e}$_{s}$, were kept in a (\textbf{e}$_{i}$,\textbf{e}$_{s}$)=(R,x) configuration for all measurements, where R represents right circular polarized light and x represents linear polarized light oriented in the ab-planes of the crystals. 

The samples were inserted into a continuous He-flow cryostat, which was horizontally mounted in the open bore of a superconducting magnet.  This experimental arrangement allowed Raman scattering measurements under the simultaneous conditions of low temperature (3-290 K) and high magnetic field (0-8 Tesla).  Field-dependent Raman measurements were performed after zero-field cooling the samples to T$\sim$3 K in order to avoid inducing the antiferromagnetic (AF) to weakly ferromagnetic (WFM) alignment of the ferromagnetic spin components in adjacent layers, which occurs for very low critical fields (H$_{c}\sim$0.15 T) in Sr$_{2}$IrO$_{4}$.\cite{Cao1998,Kim2009} 

\begin{figure}
\includegraphics[width=9cm] {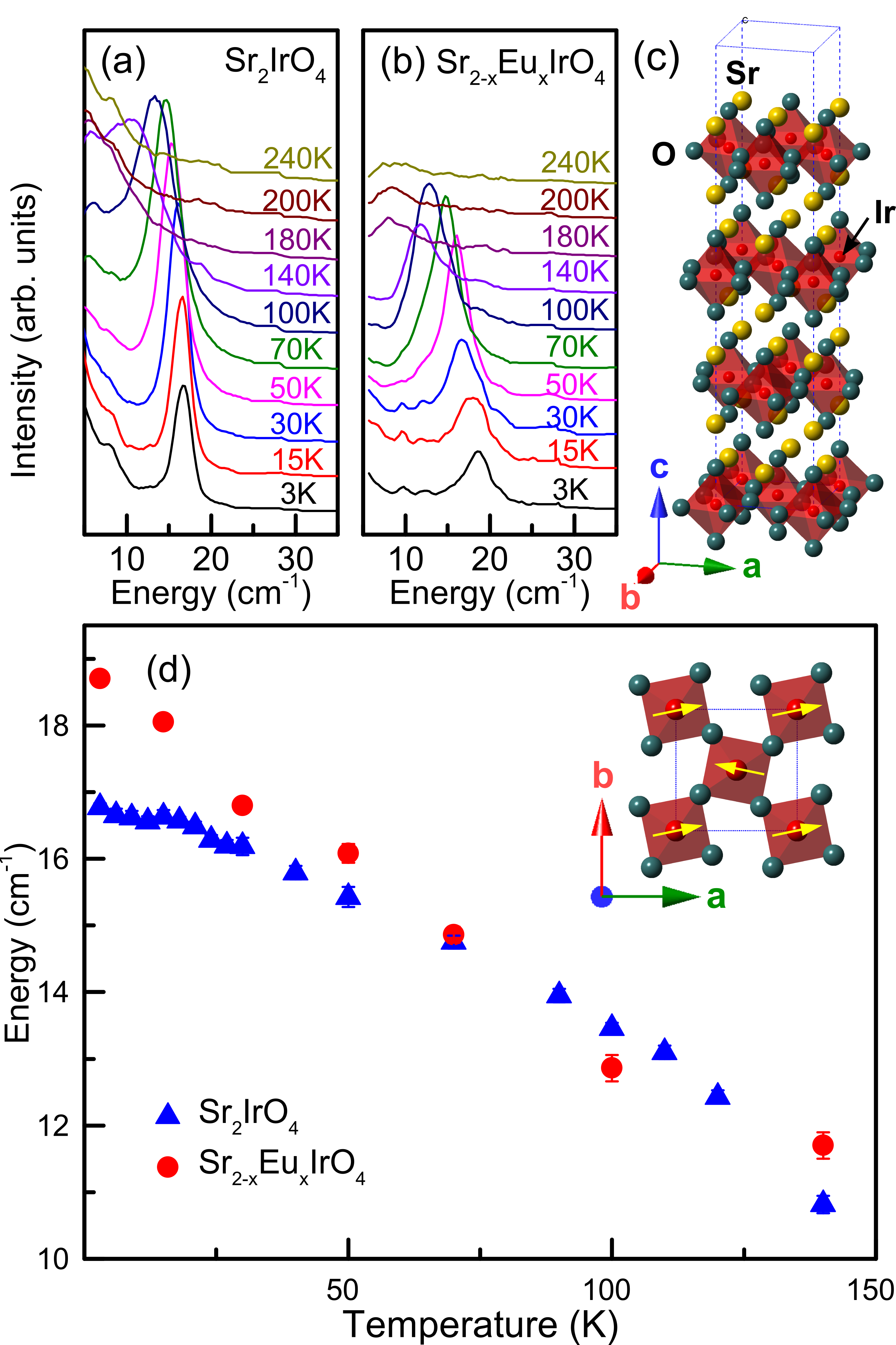}
\caption{\label{Fig1} Temperature dependence of the spin-wave spectra of (a) Sr$_{2}$IrO$_{4}$ and (b) Eu-doped Sr$_{2}$IrO$_{4}$. (c) Summary of the temperature dependence of spin-wave energies for Sr$_{2}$IrO$_{4}$ and Eu-doped Sr$_{2}$IrO$_{4}$. The left inset shows the crystal structure of Sr$_{2}$IrO$_{4}$. Because of octahedral rotations, the unit cell of Sr$_{2}$IrO$_{4}$ contains 4 layers of IrO$_{2}$. The right inset shows the in-plane orientations of the J$_{eff}$=1/2 moments. }
\end{figure}

Temperature- and field-dependent Raman scattering measurements were performed on two different Sr$_{2}$IrO$_{4}$ samples and one Eu-doped Sr$_{2}$IrO$_{4}$ sample.  The two Sr$_{2}$IrO$_{4}$ samples studied$-$one of which was used to obtain the temperature-dependent data of Fig. \ref{Fig1}(a) and the second of which was used to obtain the field-dependent data shown in Fig. \ref{Fig2}-\ref{Fig4}$-$exhibited slightly different spin-wave energies (on the order of ~1 cm$^{-1}$ or 0.13 meV energy difference).  However, the qualitative temperature- and field-dependences of the spin-wave excitation energies were nearly identical in both Sr$_{2}$IrO$_{4}$ samples. In addition to the spin-wave excitations, a temperature- and field-independent peak was observed in many of the spectra near 29 cm$^{-1}$ (peaks denoted with asterisks (*) in the H=0 T spectra of Fig. \ref{Fig2}(a) and 3(a)).  This 29 cm$^{-1}$ peak is associated with unfiltered light from the laser and was fit and subtracted from the spectra at other fields so the field-dependences of the spin-wave excitations could be more clearly observed.  Because of the very narrow linewidth of the 29 cm$^{-1}$ peak, its subtraction from the spectra did not affect our determination of the spin-wave energies at different magnetic fields.  Note that the higher frequency phonon spectra of the samples studied were also measured and the phonon results obtained were similar in most respects to results reported earlier.\cite{Cetin2012} However, the focus of this paper will be on the spin-wave excitation spectra of Sr$_{2}$IrO$_{4}$ and the phonon spectra will not be shown or discussed further here.

\section{Results}
\subsection{Temperature- and Doping-dependent Results}
\begin{figure}
\includegraphics[width=9cm] {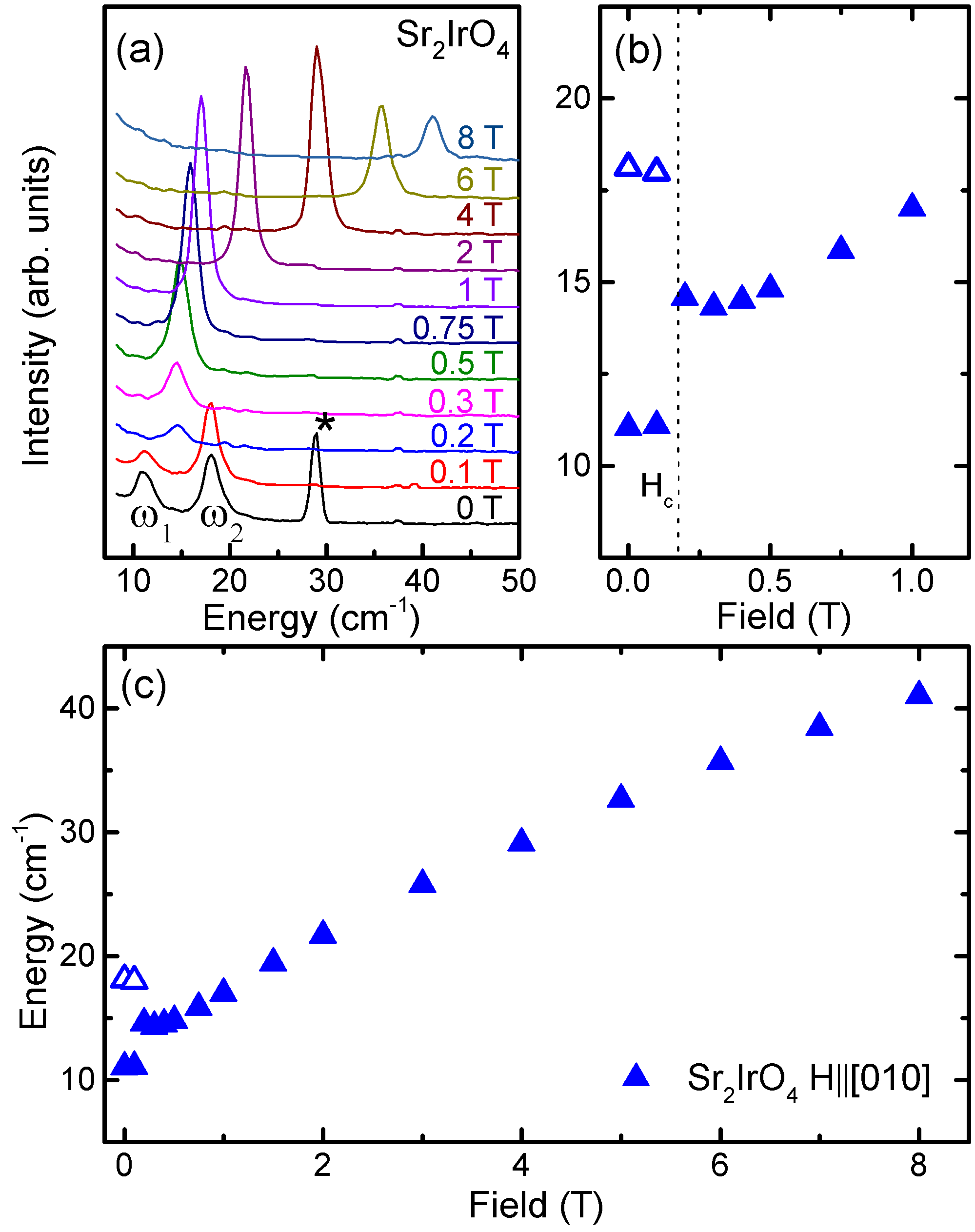}
\caption{\label{Fig2} Field dependence of spin-wave spectra of (a) Sr$_{2}$IrO$_{4}$ for H$\parallel$b-axis (=[010]) at T=3 K.  The peak marked with an asterisk (*) in the H=0 T spectra is an artifact from the laser and this peak has been removed from the spectra at other fields.  The spectra have been offset for clarity. Summary of the field-dependence of the spin wave energies for $\omega_{1}$ (filled triangles) and $\omega_{2}$ (open triangles) in Sr$_{2}$IrO$_{4}$ with H$\parallel$b-axis are shown in (b) (expanded view) and (c) (full range).}
\end{figure}

Figure \ref{Fig1}(a) shows the low frequency (5$-$35 cm$^{-1}$) excitation spectrum of Sr$_{2}$IrO$_{4}$ as a function of temperature for H=0 T.  At temperatures near T$_{N}\sim$240 K, the low energy spectrum exhibits a diffusive background, most likely associated with incoherent spin scattering.  Below T$_{N}$, this diffusive background develops into a sharp mode that increases in energy with decreasing temperature to a slightly sample-dependent value near $\omega_{2}$$\sim$17-18 cm$^{-1}$ (2.1-2.3 meV) at T=3 K. Additionally, a weak second peak near $\omega_{1}$=9-10 cm$^{-1}$ is observed in the 3 K spectrum;  this lower-energy mode is more clearly observed in the second Sr$_{2}$IrO$_{4}$ sample used for the field dependent measurements (see Fig. \ref{Fig2}) and will be discussed in more detail in the field-dependent results section below.

The effects of doping on the low energy magnetic excitation spectrum of Sr$_{2}$IrO$_{4}$ are also shown in Fig. \ref{Fig1}(b), which displays the temperature dependence of the $\sim$18 cm$^{-1}$ spin wave excitation near in Eu-doped Sr$_{2}$IrO$_{4}$.  The temperature dependences of the spin-wave energies in Eu-doped Sr$_{2}$IrO$_{4}$ (filled circles) and Sr$_{2}$IrO$_{4}$ (filled triangles) are summarized in Fig. \ref{Fig1}(c). Several slight differences between the spin wave modes in the doped and undoped Sr$_{2}$IrO$_{4}$ samples are observed:  the linewidths of the $\sim$18 cm$^{-1}$ spin wave mode are slightly broader in the doped sample compared to the undoped sample ($\Gamma_{doped}/\Gamma_{undoped}\approx$1.25), which is likely associated with greater spin and potential disorder in the doped sample.  The doped sample also exhibits a slightly higher value for the spin-wave mode energy at T=3 K, but this difference is consistent with the sample-to-sample variations we noted for the measured spin-wave energies in undoped Sr$_{2}$IrO$_{4}$; consequently, this energy difference is not believed to be significant.   Thus, the most noteworthy feature of Fig. \ref{Fig1}(b) is that there is not a substantial influence of slight doping on the \textbf{q}=0 spin-wave energies in Sr$_{2}$IrO$_{4}$.  This conclusion is consistent with evidence that electron doping in Sr$_{2}$IrO$_{4}$ causes a subtle unbuckling of the IrO$_{6}$ octahedra and a crossover to metallic behavior, but does not significantly affect the magnetic properties of Sr$_{2}$IrO$_{4}$.\cite{Ge2011}

\subsection{Field-dependent Results}
Figure \ref{Fig2}(a) shows the magnetic-field dependence of the spin-wave spectrum of Sr$_{2}$IrO$_{4}$ for a field orientation  parallel to the FM moment (i.e., H$\parallel$b-axis=[010]).  The field-dependent results for Eu-doped Sr$_{2}$IrO$_{4}$ are similar and will not be shown. Two spin-wave modes, $\omega_{1}$ and $\omega_{2}$, are clearly evident in the H=0 T spectrum at $\omega_{1}$=11 cm$^{-1}$ (1.38 meV) and $\omega_{2}$=18 cm$^{-1}$ (2.25 meV).  As a first step towards identifying these modes, note that in the simplest description of Sr$_{2}$IrO$_{4}$ as a two-dimensional canted antiferromagnet$-$which ignores, in particular, interlayer coupling between the antiferromagnetically coupled layers \cite{Thio1988}$-$the two-fold degenerate \textbf{q}=0 spin-wave branch is expected to split into a low-frequency ``ferromagnetic (FM) mode" and a higher frequency ``antiferromagnetic (AF) mode", associated with precession of the spins about the FM and AF axes, respectively.\cite{White1982,Cottam1986,Gozar2004,Bahr2014}  However, we can likely rule out assigning either $\omega_{1}$ or $\omega_{2}$ to the FM mode of Sr$_{2}$IrO$_{4}$, because previous ESR measurements have reported that the FM spin-wave mode in Sr$_{2}$IrO$_{4}$ has an H$\approx$0 T value of $\omega_{FM}$=0.32 cm$^{-1}$,\cite{Bahr2014} which is well below the spectral range of our light scattering study.  We can also rule out the possibility that the modes at $\omega_{1}$ and $\omega_{2}$ in Fig. \ref{Fig1}(a) are the same spin-wave mode associated with different magnetic domains in Sr$_{2}$IrO$_{4}$.  Magnetic domains have been reported in Sr$_{2}$IrO$_{4}$, but likely involve simple 90$^{\circ}$ rotations of the unit cell,\cite{Dhital2013} which cannot account for the significantly different energies (~$\sim$1 meV) of the $\omega_{1}$ and $\omega_{2}$ modes shown in Sr$_{2}$IrO$_{4}$ (see Fig. \ref{Fig2}(a)).  Magnetic domains with a different stacking sequence of the layers$-$such as domains already in the WFM phase at H=0 T$-$would cause the same spin-wave mode to have slightly different energies in the different domains.  However, the energy difference in this case would probably not be large enough to account for the large ($\sim$ 1 meV) observed energy difference between the $\omega_{1}$ and $\omega_{2}$ spin wave modes in Sr$_{2}$IrO$_{4}$. Further, to our knowledge there have been no reports that domains associated with the WFM phase are present at H=0 T in Sr$_{2}$IrO$_{4}$. 

Therefore, the two spin-wave modes $\omega_{1}$ and $\omega_{2}$ in Fig. \ref{Fig2}(a) are most likely associated with the effects of interlayer coupling between antiferromagnetically coupled IrO layers.  As discussed by Thio et al. for La$_{2}$CuO$_{4}$,\cite{Thio1990} interlayer coupling between the two inequivalent (antiferromagnetically coupled) layers in Sr$_{2}$IrO$_{4}$ results in a magnetic unit cell that contains four spins and two 2-fold FM and AF magnon branches whose degeneracies at H=0 T are split by interlayer coupling.\cite{Chovan2000}  We associate the spin-wave modes $\omega_{1}$ and $\omega_{2}$ in Sr$_{2}$IrO$_{4}$ with the in-phase and out-of-phase combinations of the AF spin-waves on adjacent layers, respectively. This interpretation is supported by the observed reduction from two \textbf{q}=0 AF spin-wave modes in the antiferromagnetic (AF) phase of Sr$_{2}$IrO$_{4}$$-$which has two magnetically inequivalent layers per unit cell in the simplest model description$-$to a single \textbf{q}=0 AF spin-wave mode (see Figs. \ref{Fig2}(a), \ref{Fig2}(b), and \ref{Fig2}(c)) in the weakly ferromagnetic (WFM) phase of Sr$_{2}$IrO$_{4}$, which has only a single layer per unit cell.  In particular, in the WFM phase, the out-of-phase AF spin-wave mode $\omega_{2}$ becomes a zone-boundary mode and only the in-phase AF mode $\omega_{1}$ is expected to be present at \textbf{q}=0.  

The importance of interlayer coupling on the spin-wave excitation spectrum of Sr$_{2}$IrO$_{4}$ is also supported by the abrupt increase in the in-phase AF spin-wave energy ( $\omega_{1}$=3.4 cm$^{-1}$ or 0.43 meV) (Figs. \ref{Fig2}(b) and \ref{Fig2}(c)) at H$_{c}$, which reflects an increase in the AF spin-wave stiffness through the AF-to-WFM transition.  The energy shift of $\omega_{1}$ at H$_{c}$ allows an estimate of the interlayer coupling energy in Sr$_{2}$IrO$_{4}$:  Using the measured change in the energy of $\omega_{1}$ at H$_{c}$ (see Fig. \ref{Fig2}(b)) and the result that,\cite{Chovan2000, Benfatto2006b}

\begin{equation}
4JJ_{\perp}=[\omega_{1}^{2}(H_{c}^{+})-\omega_{1}^{2}(H_{c}^{-})]/\sqrt{2}
\label{eq0}
\end{equation}
we find $4JJ_{\perp}$$\sim$59 cm$^{-2}$ in Sr$_{2}$IrO$_{4}$, giving an estimate for the value of the interlayer coupling energy J$_{\perp}$$\sim$0.018 cm$^{-1}$ (2.3 $\mu$eV) (using J$\sim$800 cm$^{-1}$).\cite{Fujiyama2012} This estimate of J$_{\perp}$ is consistent with published reports for Sr$_{2}$IrO$_{4}$, including estimates based upon the measured critical field H$_{c}$ in Sr$_{2}$IrO$_{4}$: J$_{\perp}$=mH$_{c}$/S$^{2}$,\cite{Thio1988, Chovan2000} which gives J$_{\perp}$$\sim$3 $\mu$eV for Sr$_{2}$IrO$_{4}$, using m=0.07$\mu_{B}$ per Ir atom, H$_{c}$=0.15 T, and S=1/2.

\begin{figure}
\includegraphics[width=9cm] {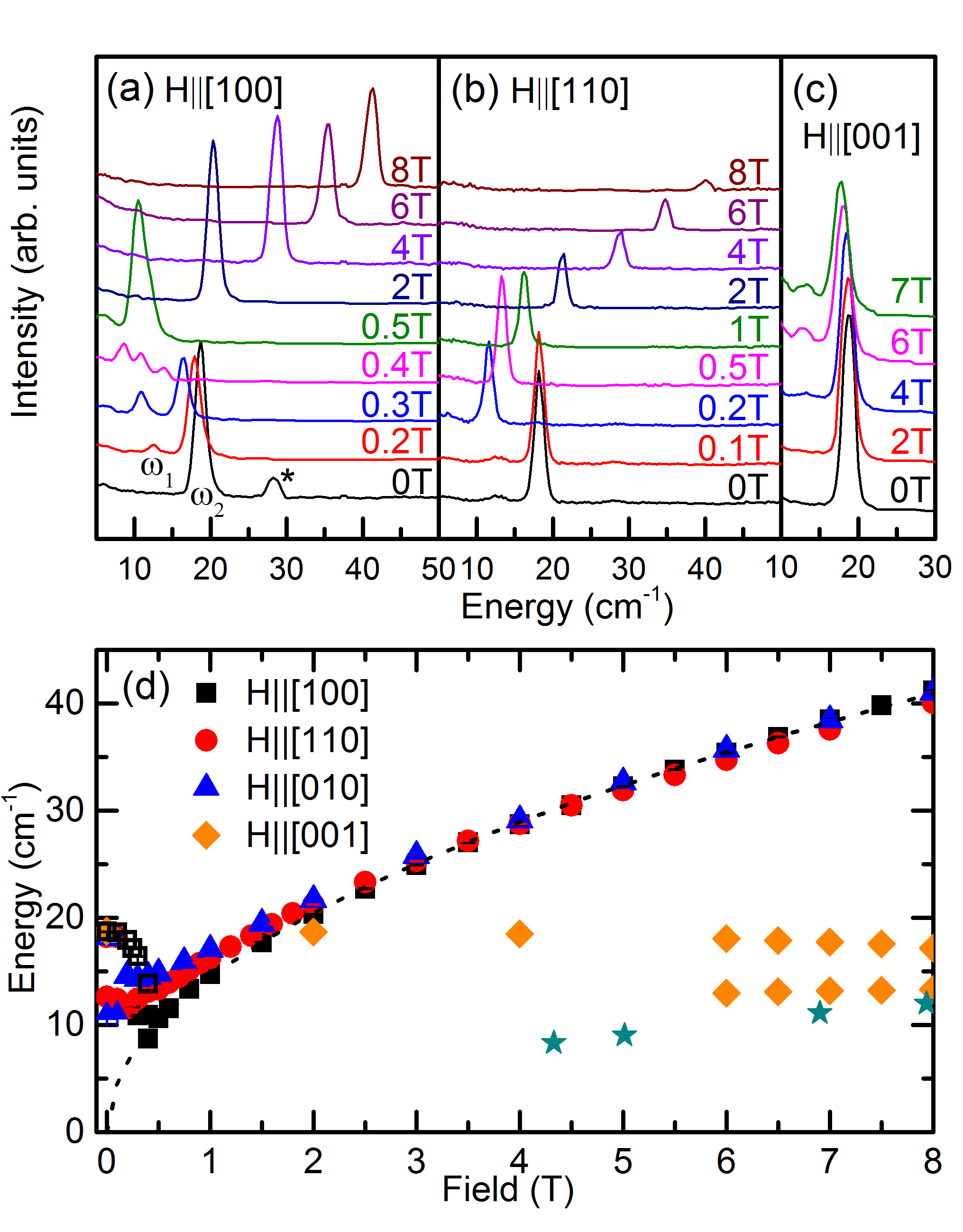}
\caption{\label{Fig3} Field dependences of the spin-wave spectra of Sr2IrO4 for (a) H$\parallel$[100], (b) H$\parallel$[110], and (c) H$\parallel$[001] at T=3 K.  The peak marked with an asterisk (*) in the H=0 T spectrum is an artifact from the laser and has been removed from the spectra at other fields.  The spectra have been offset for clarity.  (d) Summary of the field-dependences of the spin wave energies for different applied field orientations for both $\omega_{1}$(filled symbols) and $\omega_{2}$(open symbols). Also shown for comparison are results from ESR measurements \cite{Bahr2014} for H$\parallel$[001] (filled stars).  The dashed line is a fit to the data with the functional form $\omega$=$\sqrt{\gamma H}$ using $\gamma$=209.38 cm$^{-2}$T$^{-1}$.}
\end{figure}

The magnetic-field dependences of the AF spin-wave energies $\omega_{1}$ and $\omega_{2}$ of Sr$_{2}$IrO$_{4}$ are shown for different applied field orientations in Fig. \ref{Fig3}.  Figure \ref{Fig3}(a) shows the magnetic field dependences of $\omega_{1}$ and $\omega_{2}$ with H roughly parallel to the spin direction, H$\parallel$a-axis=[100], while Fig. \ref{Fig3}(b) shows the magnetic field dependences of $\omega_{1}$ and $\omega_{2}$ with H oriented roughly 45$^{\circ}$ from the a-axis, i.e., H$\parallel$[110]. Note that the H=0.4 T spectrum in Fig. 3 (a) shows three peaks, consisting of a superposition between the two spin-wave modes of the AF phase and the single spin-wave mode in the WFM phase.  This superposition is consistent with a coexistence of AF and WFM phases expected near the first-order transition at H$_{c}$.

Also shown in Fig. \ref{Fig3}(c) is the magnetic field dependence of the \textbf{q}=0 spin-wave spectrum in Sr$_{2}$IrO$_{4}$ for the out-of-plane magnetic field orientation, i.e., with H roughly parallel to the c-axis direction, H$\parallel$[001].  The $\omega_{1}$$\sim$8 cm$^{-1}$ (not shown) and $\omega_{2}$$\sim$18 cm$^{-1}$ spin-wave mode energy exhibits a much weaker magnetic field dependence for H$\parallel$c-axis=[001], consistent with previous electron spin resonance (ESR) results (filled stars).\cite{Bahr2014}  An additional weak mode develops near $\sim$13 cm $^{-1}$ for H$>$4 T with H$\parallel$c-axis=[001]. This mode may be associated with the presence of a small in-plane field caused by a slight misalignment of the magnetic field in the H$\parallel$c-axis configuration, which can induce an AF-to-WFM transition$-$and a lower value for the spin-wave energy (as discussed above)$-$in parts of the sample.

\begin{figure}[!]
\includegraphics {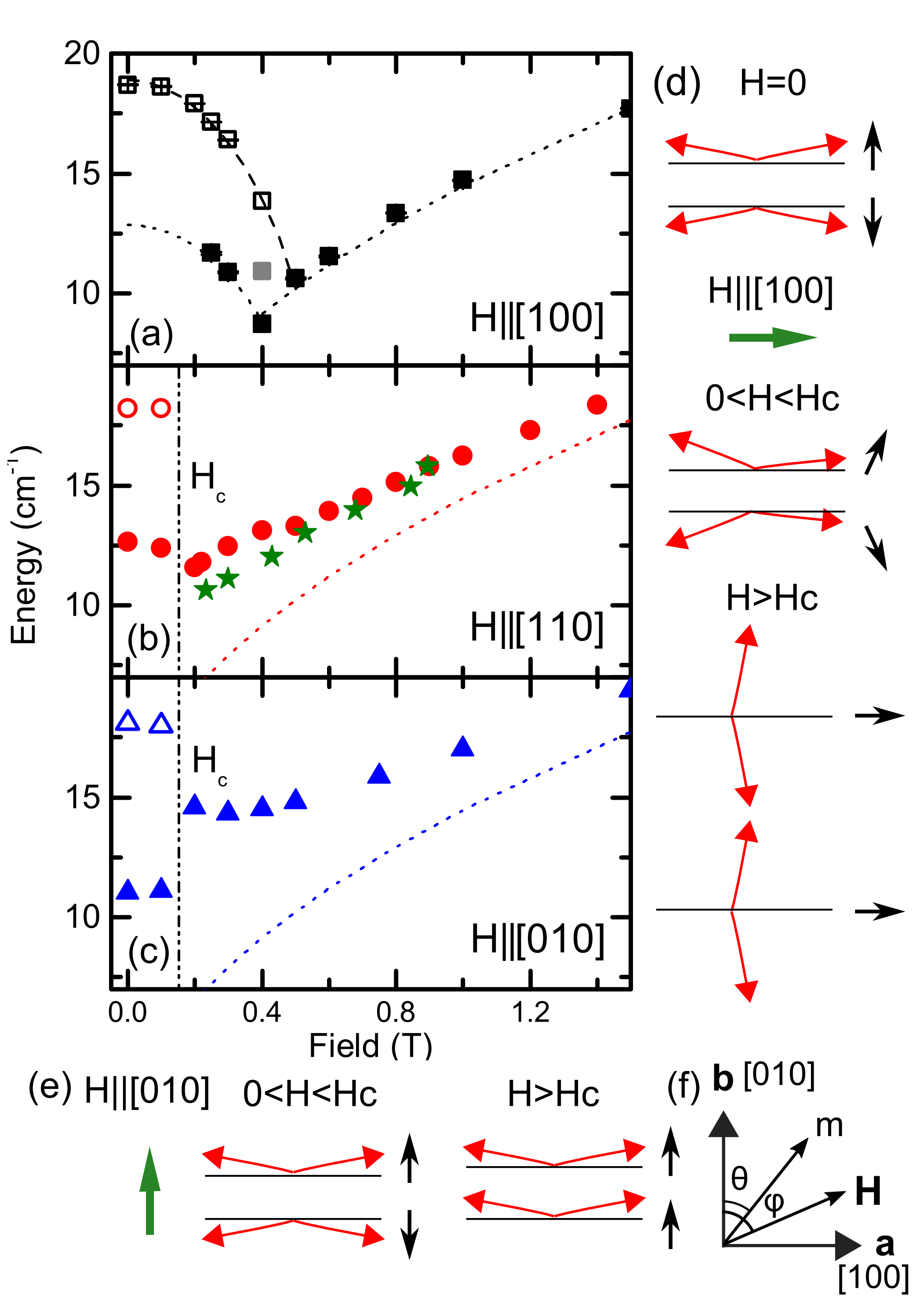}
\caption{\label{Fig4} Summaries of the field dependences of the spin-wave energies of Sr$_{2}$IrO$_{4}$ for different in-plane field orientations, including (a) H$\parallel$a-axis=[100], (b) H$\parallel$[110], and (c) H$\parallel$b-axis=[010] at T=3 K. (Closed symbols= $\omega_{1}$ mode, Open symbols= $\omega_{2}$ mode) The dashed lines are plots of $\omega$=$\sqrt{\gamma H}$ with $\gamma$=209.38 cm$^{-2}$T$^{-1}$ for comparison with the data.  The dotted-dashed line is a fit to the data with the functional form of $\omega$=$\sqrt{\Delta^{2}-\alpha H^{2}}$ with $\Delta$=18.9 cm$^{-1}$ and $\alpha$=1014 cm$^{-2}$T$^{-2}$.  Also shown for comparison are results from ESR measurements\cite{Bahr2014} for H$\parallel$[110] (filled stars).  (d) Schematic illustration of the rotation of the staggered spin components (red arrows) and uniform spin component (black arrows) on adjacent layers for an applied field (green arrow) oriented transverse to easy axis direction of FM component of the spin, H$\parallel$a-axis=[100], illustrating the continuous rotation of the spins on adjacent layers for this applied field orientation.  (e)  Schematic illustration of the rotation of the staggered spin components (red arrows) and uniform spin component (black arrows) on adjacent layers for an applied field (green arrow) oriented parallel to easy axis direction of FM component of the spin, H$\parallel$ b-axis=[010], illustrating the abrupt flipping of the spins in one layer for this field orientation. (f) Diagram showing the angle $\theta$ of the FM component of the spins (\textbf{m}) and the angle $\phi$ of the applied field (H) relative to the easy axis in-plane orientation of \textbf{m} (i.e., [010]).}
\end{figure}

\section{Discussion}

The central result of this study concerns the magnetic-field-dependences (0$\leq$H$\leq$8 T) of the AF spin-wave mode energies summarized in Fig. \ref{Fig3}(d) for different in-plane magnetic field orientations, H$\parallel$a-axis=[100] (filled squares), H$\parallel$b-axis=[010] (filled triangles), and H$\parallel$[110] (filled circles).  Fig. \ref{Fig3}(d) illustrates that there are 2 distinct field regimes for the in-plane spin dynamics in Sr$_{2}$IrO$_{4}$, (A) an isotropic regime for H$\gtrsim$1.5T and (B) an anisotropic regime for H$\lesssim$1.5T.

\subsection{Isotropic regime H$\gtrsim$1.5T }
For H$>$1.5 T, the in-plane spin dynamics are isotropic and the AF spin-wave mode $\omega_{1}$ energy in the WFM phase region is well-described by a square-root field dependence, $\omega_{1}=\sqrt{\gamma H}$, with $\gamma$=209.4 cm$^{-2}$T$^{-1}$ (dashed line).  The isotropic square-root field dependence for H$>$1.5 T indicates that the FM components of the spins simply follow the applied field direction in Sr$_{2}$IrO$_{4}$, due to the dominant interaction between the applied field and the weak FM moments induced by the DM interaction.\cite{Cottam1986,Benfatto2006a,Bahr2014}

The spin dynamics above H$>$1.5 T in Sr$_{2}$IrO$_{4}$ are consistent with an isotropic, two-dimensional effective S=1/2 Hamiltonian given by: \cite{Jackeli2009,Bahr2014}
\begin{equation}
H_{12}=J\vec{S_{1}}\cdot\vec{S_{2}}+\Gamma S_{1}^{z}S_{2}^{z}+D(S_{1}^{x}S_{2}^{y}-S_{1}^{y}S_{2}^{x})
\label{eq1}
\end{equation}
where the first term (J) is associated with isotropic antiferromagnetic exchange between the two inequivalent spins, 1 and 2, in the IrO plane, the second term ($\Gamma$) represents symmetric exchange anisotropy that favors collinear c-axis spin order, and the third term (D) represents antisymmetric exchange anisotropy that favors canted in-plane spin order.  Bahr \textit{et al}. predict that for $\Gamma$,D $\ll$ J, the AF spin-wave energy associated with the model Hamiltonian in Eq.(\ref{eq1}) should have a field-dependence given by,\cite{Bahr2014}	
\begin{equation}
\omega_{1}\approx\sqrt{\Delta^{2}+8Jm_{FM}H}
\label{eq2}
\end{equation}
where J$\sim$100 meV in Sr$_{2}$IrO$_{4}$ and m$_{FM}$ is the FM canting moment. This prediction is consistent with the square-root field dependence we observe for $\omega_{1}$ in Fig. \ref{Fig3}(d). Using our value of $\gamma$=209.4 cm$^{-2}$T$^{-1}$ from the fit to the data in Fig. \ref{Fig3}(d) (dashed line) with $\Delta$$\sim$ 0, we obtain an estimated FM canting moment of m$_{FM}$$\sim$$\gamma$/8J $\sim$ 0.07$\mu_{B}$ in Sr$_{2}$IrO$_{4}$, which is consistent with other estimates (e.g., see ref. 16).  Notably, the $\gamma$ value determined from the field-dependence of the AF spin wave in La$_{2}$CuO$_{4}$ ($\gamma_{LCO}$= 22.6 cm$^{-2}T^{-1}$)\cite{Gozar2004} is much smaller than our value for Sr$_{2}$IrO$_{4}$, reflecting the much smaller FM moment associated with spin canting in La$_{2}$CuO$_{4}$ (m$_{FM}$$\sim$0.002$\mu_{B}$).\cite{Thio1988}

\subsection{Anisotropic regime H$\lesssim$1.5T}
Figure \ref{Fig3}(d) shows that the field-dependent spin-wave dynamics for H$<$1.5 T are highly anisotropic in the planes, revealing interaction effects in Sr$_{2}$IrO$_{4}$ that are not accounted for in Eq.(\ref{eq1}).  A more detailed view of the anisotropic magnetic field dependence of the AF spin wave energy in Sr$_{2}$IrO$_{4}$ is provided in Fig. \ref{Fig4}, which shows the field-dependences of spin-wave energies $\omega_{1}$ and $\omega_{2}$ in the field range 0$\leq$H$\leq$1.5 T for several in-plane field orientations, including (a) H$\parallel$a-axis=[100], (b) H$\parallel$[110], and (c) H$\parallel$b-axis=[010]. As discussed above, Fig. \ref{Fig4}(c) shows that the in-phase AF spin-wave energy $\omega_{1}$ exhibits an abrupt increase in energy ($\Delta$$\omega_{1}$=3.4 cm$^{-1}$ or 0.43 meV) through the AF-to-WFM spin-flop transition at H$_{c}$$\sim$0.15 T when the applied field is oriented in the direction of the FM (uniform) spin component of the spins, H$\parallel$b-axis=[010].  This behavior indicates that the AF-to-WFM transition in Sr$_{2}$IrO$_{4}$ occurs via a discontinuous spin-flop transition, and results in a discontinuous change in interlayer coupling, when the applied field is oriented along the weak FM component of the spins (see Fig. \ref{Fig4}(e)).

On the other hand, Fig. \ref{Fig4}(a) shows that when the applied field is oriented parallel to the staggered spins, H$\parallel$a-axis=[100], AF spin-wave modes $\omega_{1}$ and $\omega_{2}$ exhibit ``soft mode" behavior: the field-dependence of $\omega_{2}$ with H$\parallel$a-axis=[100] is well described by the functional form $\omega_{2}=\sqrt{\Delta^{2}-\alpha H^{2}}$ (dashed-dotted line) with $\Delta$=18 cm$^{-1}$ and $\alpha$=1014 cm$^{-2}$$T^{-2}$. The soft spin-wave mode behavior shown in Fig. \ref{Fig4}(a) indicates that the AF-to-WFM transition involves a continuous spin reorientation and a gradual crossover when H$\parallel$a-axis=[100] (Figure \ref{Fig4}(d)). The field-dependence of the AF spin wave energy of Sr$_{2}$IrO$_{4}$ for H$\parallel$[110], shown in Fig. \ref{Fig4}(b), exhibits behavior intermediate to that observed for the H$\parallel$[010] and H$\parallel$[100] orientations.  Also shown for comparison in Fig. \ref{Fig4}(b) is the field-dependence of the AF spin-wave mode $\omega_{AF}$ determined from electron-spin-resonance (ESR) measurements with H$\parallel$[110] (filled stars),\cite{Bahr2014} showing that there is a good agreement between the AF spin-wave energies measured with Raman scattering and ESR for this H$\parallel$[110] orientation.  

The dramatic difference in the nature of the AF-to-WFM transition for different in-plane field orientations (Fig. \ref{Fig4}) reflects the importance of in-plane anisotropy for H$\lesssim$1.5T in Sr$_{2}$IrO$_{4}$. Similar effects of in-plane anisotropy on the spin dynamics of ferrimagnets \cite{Horner1968} and canted antiferromagnets \cite{Pincus1960,Shapiro1974,Fainstein1993} have been observed previously, particularly in iridates,\cite{Wang2014} cuprates \cite{Thio1990,Fainstein1993} and ferrites. \cite{Pincus1960, Levinson1969,Koshizuka1988,Gorodetsky1970,Shapiro1974} The AF spin-wave mode softening observed in Sr$_{2}$IrO$_{4}$ (Fig. \ref{Fig4}(a)) reflects a continuous decrease in the interlayer exchange energy in Sr$_{2}$IrO$_{4}$ with applied field for H$\perp$m$_{FM}$, caused by the continuous field-induced rotation of the FM moments in opposite directions in the antiferromagnetically coupled layers (see Fig. \ref{Fig4}(d)). 

An estimate of the in-plane anisotropy field, H$_{A}$, can be obtained from our data by first developing a simple phenomenological description of the interlayer coupling energy between two adjacent layers can be written, E$_{\perp}$$\sim$J$_{\perp}$cos(2$\theta$)=J$_{\perp}$(1-2sin$^{2}$($\theta$)), where $\theta$ is the angle between the FM spin components and their zero-field (easy axis) directions in the each of two layers (see Fig. \ref{Fig4}(f)).  The interlayer coupling energy can be written in terms of the applied in-plane field H, using the result that the equilibrium in-plane orientation for the weak FM moment in each layer for a particular field H is given by:\cite{Fainstein1993}
\begin{equation}
(H_{DM}/H_{E})Hsin(\phi-\theta)=(H_{A})sin(2\theta)
\label{eq3}
\end{equation}
where H$_{DM}$ is the Dzyaloshinskii-Moriya field, H$_{E}$ is the exchange field, H$_{A}$ is the in-plane anisotropy field, H is the applied field, $\theta$ is the angle between the FM spin component, m$_{FM}$, and its zero-field (easy axis) orientation, $\phi$ is the angle between the applied field and the easy-axis, and assuming H$_{E}$$\gg$H$_{DM}$$\gg$H$_{A}$$\sim$H.  Equation (\ref{eq3}) shows that a field applied perpendicular to the easy-axis orientation of m$_{FM}$ (i.e., H$\perp$m$_{FM}$ or $\phi$=$\pi$/2), which is the field orientation for which we observe soft magnon behavior (see Fig. \ref{Fig4}(a)), induces an in-plane rotation of m$_{FM}$ by an angle $\theta$ that increases continuously with the applied field according to  sin($\theta$)=(H$_{DM}$/2H$_{A}$H$_{E}$)H, as schematically depicted in Fig. \ref{Fig4}(d).  In the AF phase, the ferromagnetic components in adjacent layers rotate in opposite directions in response to an applied transverse in-plane field. Consequently, the interplane exchange energy for a pair of coupled layers will continuously decrease with field H for H$<$H$_{c}$ according to E$_{\perp}$$\sim$J$_{\perp}$(1-2$\beta^{2}$H$^{2}$), where $\beta$=(H$_{DM}$/2H$_{E}$H$_{A}$).  This functional form for E$_{\perp}$ is consistent with the observed field-dependences of the spin wave mode energies near H$\sim$0.4 T for H$\parallel$a-axis=[100] (see Fig. \ref{Fig4}(a)). Note that the interplane exchange energy goes to zero, E$_{\perp}\rightarrow$0, at a critical field given by  H$_{c}$=$\sqrt{2}$(H$_{E}$/H$_{DM}$)H$_{A}$. Using our rough measurement of the field at which the AF spin-wave mode energy approaches zero, H$_{c}$$\approx$0.4T, and an estimate of the ratio (H$_{DM}$/H$_{E}$) using tan(2$\xi$)=(H$_{DM}$/H$_{E}$),\cite{Bahr2014} where $\xi$=11$^{\circ}$ is the canting angle, we obtain a value for the in-plane anisotropy field in Sr$_{2}$IrO$_{4}$, H$_{A}$=1/$\sqrt{2}$(H$_{DM}$/H$_{E}$)H$_{c}$$\approx$0.1T. This estimate compares well with the coercive field $\sim$0.15T needed to induce an abrupt ``spin-flip" transition between AF and WFM phases for H$\parallel$b-axis =[010] (see Fig. \ref{Fig4}(c)). Additionally, the minimum value for the spin-wave energy at H=0.4 T (see Fig. \ref{Fig4}(a)), $\Delta$$\sim$8 cm$^{-1}$ (1 meV), offers a good estimate of the spin-gap energy in Sr$_{2}$IrO$_{4}$ without the effects of interlayer coupling. 

\section{Summary}
In the field- and temperature-dependent Raman scattering studies of the angle-dependence of spin excitations of Sr$_{2}$IrO$_{4}$ presented here, we show clear evidence for a magnetic field scale H$\sim$1.5 T above which the in-plane spin dynamics behave in accordance with the predictions of an isotropic, two-dimensional effective S=1/2 Hamiltonian.  The field-dependence of spin-wave excitations in this ``high field" regime are isotropic, two-dimensional, and solely governed by the interplay between the applied field and the FM component of the spins associated with the DM interaction.  However, dramatic deviations from this isotropic and two-dimensional behavior are clearly observed at lower fields, H$<$1.5 T, manifested, for example, in a highly anisotropic field-dependence of the spin dynamics and interlayer-exchange-split spin-wave modes.  Particularly noteworthy is the observation of field-induced magnon soft mode behavior near H$_{c}$ for a field applied transverse to the FM spin components, H$\perp$m$_{FM}$, which reveals a continuous spin rearrangement transition at the antiferromagnetic-to-weakly ferromagnetic transition at H$_{c}$ in Sr$_{2}$IrO$_{4}$. Our results also show that when the in-plane field is aligned perpendicular to the easy-axis direction of the FM moment, the field dependence of the \textbf{q}=0 spin-wave energy evolves according to $\omega$$\sim$$H^{1/2}$ above H$_{c}$, i.e., in a manner consistent with a 2D canted antiferromagnet with no spin gap. These studies highlight the importance of considering in-plane anisotropy and interlayer coupling effects on the low energy spin dynamics when interpreting and calculating the low-field magnetic and dynamical properties of Sr$_{2}$IrO$_{4}$.\\

\textbf{Acknowledgments}$-$Research was supported by the National Science Foundation under Grant NSF DMR 14-64090. Work at the University of Kentucky was supported by the National Science Foundation via Grant Nos. DMR-1265162. Work at Argonne National Laboratory (crystal growth and magnetic characterization) was supported by the U.S. Department of Energy Office of Science, Basic Energy Sciences, Materials Science and Engineering Division. We thank Sam Gleason for useful discussions

\end{document}